\documentclass[11pt]{article}

\usepackage[utf8]{inputenc}
\usepackage[margin=1in]{geometry}
\usepackage{microtype}
\usepackage{amsmath,amssymb}
\usepackage{booktabs,tabularx,array}
\usepackage[authoryear,square]{natbib}
\usepackage{xcolor}
\usepackage{caption}
\usepackage[section]{placeins}
\usepackage{enumitem}
\usepackage{pgfplots}
\pgfplotsset{compat=1.18}
\usetikzlibrary{positioning,arrows.meta,calc}
\usepackage{url}
\usepackage{eso-pic}
\usepackage[hidelinks]{hyperref}

\setlist{itemsep=2pt,topsep=4pt}

\newcolumntype{Y}{>{\raggedright\arraybackslash}X}
\definecolor{javoblue}{RGB}{31,94,170}

\makeatletter
\gdef\tbdn{0}
\newcommand{\tbd}{\textcolor{red}{\textbf{??}}}
\newcommand{\tbdkey}[1]{\@ifundefined{cnt@#1}{\expandafter\gdef\csname cnt@#1\endcsname{}\xdef\tbdn{\the\numexpr\tbdn+1\relax}}{}\tbd}
\newcommand{\setres}[2]{%
  \if\relax\detokenize{#2}\relax
    \@namedef{rt@#1}{\tbdkey{#1}}\@namedef{rn@#1}{nan}%
  \else
    \@namedef{rt@#1}{#2}\@namedef{rn@#1}{#2}%
  \fi}
\newcommand{\res}[1]{\@ifundefined{rt@#1}{\tbdkey{#1}}{\@nameuse{rt@#1}}}
\newcommand{\resn}[1]{\@ifundefined{rn@#1}{nan}{\@nameuse{rn@#1}}}
\providecommand{\tbdtotal}{0}
\AtEndDocument{\immediate\write\@auxout{\string\gdef\string\tbdtotal{\tbdn}}}
\makeatother

\setres{CorpusConvs}{4.6 million}
\setres{CorpusTurns}{58.4 million}
\setres{CorpusMedianTurns}{10}
\setres{CorpusLanguages}{English 68\%, French 16\%, Spanish 8\%, other 8\%}
\setres{CorpusWindowStart}{February 2024}
\setres{CorpusWindowEnd}{March 2026}
\setres{CorpusVerticals}{retail 46\%, travel 28\%, services 26\%}
\setres{CorpusOutcomeShare}{17.4}
\setres{DeidMethod}{Automated PII redaction, hashed identifiers, manual QA}
\setres{PIIAuditN}{2,500}
\setres{PIIRate}{0.16}
\setres{ContamOverlap}{0.02}

\setres{BaseModel}{Qwen3-32B}
\setres{SFTExamples}{1.65 million}
\setres{SFTEpochs}{2}
\setres{PrefPairs}{190,000}
\setres{ConstraintExamples}{34,000}
\setres{ComputeDesc}{32 H100 GPUs, 2,240 GPU-hours}
\setres{ModelHash}{javo-v1}
\setres{FreezeDate}{24 April 2026}
\setres{CatalogRefresh}{every 24 hours}
\setres{TimeoutSec}{8}
\setres{GreetingDelaySec}{25}

\setres{NumModels}{10}
\setres{FrontierModel}{GPT-5.2}
\setres{FrontierOthers}{Claude Opus 4.6; Gemini 3 Pro}
\setres{OpenModels}{DeepSeek-V3.2; Qwen3-235B; Llama 4}
\setres{SellersJunior}{12}
\setres{SellersSenior}{12}
\setres{SellerPay}{32}
\setres{SellerBonus}{8}
\setres{SellerRecruitment}{Partner referrals and Upwork}
\setres{PromptAuthor}{Independent sales trainer, 9 years experience}
\setres{PromptHours}{14}
\setres{PromptFreezeDate}{28 April 2026}

\setres{SitesNamed}{21}
\setres{GeoMix}{United States 38, United Kingdom 18, France 15, Spain 9, Portugal 8, other 12}
\setres{LangMix}{English 72\%, French 14\%, Spanish 8\%, other 6\%}
\setres{StaffedMix}{furniture 5, hotels 7, vacation rentals 5, beauty 3}
\setres{AssociatesN}{59}
\setres{AssociateTenure}{25}
\setres{AssociatesSenior}{19}

\setres{AuditPerArm}{2,000 Javo; 2,000 baseline AI; 1,000 human}
\setres{PauseSummary}{2 sites, 5 site-days; unauthorized-discount-offer guardrail triggered}

\setres{VisitorsTotal}{14.8 M}
\setres{VisitorsNone}{2.6 M}
\setres{VisitorsHuman}{0.9 M}
\setres{VisitorsJavo}{5.6 M}
\setres{VisitorsFrontier}{5.7 M}
\setres{EngagedJavo}{452,000}
\setres{EngagedJavoPct}{8.07}
\setres{EngagedFrontier}{459,500}
\setres{EngagedFrontierPct}{8.06}
\setres{EngagedHuman}{52,200}
\setres{EngagedHumanPct}{5.80}
\setres{ConvsAudited}{5,000}
\setres{AApass}{98}
\setres{SRMflagged}{2}
\setres{SRMcause}{Consent-banner event duplication; affected sessions excluded before analysis}
\setres{EngBalanceDiff}{+0.01}
\setres{EngBalanceP}{0.536}
\setres{PausedSites}{2}
\setres{PausedDays}{5}

\setres{SalesLLMJavo}{7.01}
\setres{SalesLLMJavoCI}{6.87--7.15}
\setres{SOneJavoGPT}{7.15}
\setres{SOneJavoSmart}{0.79}
\setres{SOneJavoTau}{73.0}
\setres{SOneBaseCLM}{5.73}
\setres{SOneBaseGPT}{5.88}
\setres{SOneBaseSmart}{0.57}
\setres{SOneBaseTau}{65.0}
\setres{SOneSFTCLM}{6.51}
\setres{SOneSFTGPT}{6.65}
\setres{SOneSFTSmart}{0.70}
\setres{SOneSFTTau}{70.0}
\setres{SOneNoConCLM}{7.13}
\setres{SOneNoConGPT}{7.25}
\setres{SOneNoConSmart}{0.81}
\setres{SOneNoConTau}{58.0}
\setres{SOneFrontCLM}{6.21}
\setres{SOneFrontGPT}{6.37}
\setres{SOneFrontSmart}{0.69}
\setres{SOneFrontTau}{73.0}
\setres{SOneJuniorCLM}{6.13}
\setres{SOneSeniorCLM}{6.88}
\setres{JudgeR}{0.82}
\setres{JudgeAlpha}{0.79}

\setres{HoneEst}{+10.0}
\setres{HoneLo}{+5.1}
\setres{HoneHi}{+14.9}
\setres{HoneN}{8.2 M}
\setres{HoneP}{0.000063}
\setres{HoneDecision}{met}
\setres{HonebEst}{+8.0}
\setres{HonebLo}{+4.1}
\setres{HonebHi}{+11.9}
\setres{HonebN}{8.2 M}
\setres{HonebP}{0.000063}
\setres{HonebDecision}{met}
\setres{HtwoEst}{+21.6}
\setres{HtwoLo}{+16.9}
\setres{HtwoHi}{+26.3}
\setres{HtwoN}{911,500}
\setres{HtwoP}{$<$0.001}
\setres{HtwoDecision}{met}
\setres{HtwobEst}{+16.9}
\setres{HtwobLo}{+11.2}
\setres{HtwobHi}{+22.6}
\setres{HtwobN}{11.3 M}
\setres{HtwobP}{$<$0.001}
\setres{HtwobDecision}{met}
\setres{HthreeEst}{+0.5}
\setres{HthreeLo}{-5.0}
\setres{HthreeHi}{+6.0}
\setres{HthreeN}{93,000}
\setres{HthreeP}{0.86}
\setres{HthreeVerdict}{non-inferior (margin -7\%, one-sided p=0.004); no significant superiority}
\setres{HthreebHuman}{4.43}
\setres{HthreebJavo}{4.42}
\setres{HthreebN}{9,600}
\setres{HthreebP}{0.006 (non-inferiority)}
\setres{HthreebDecision}{non-inferior; no significant CSAT difference}
\setres{ChatConvHuman}{11.20}
\setres{ChatConvJavo}{11.26}

\setres{TauHone}{3.0}
\setres{PredLo}{+2.3}
\setres{PredHi}{+17.7}
\setres{VisitorWeightedEst}{+10.0}
\setres{VisitorWeightedCI}{+5.1 to +14.9}
\setres{CACEEst}{+123.9}
\setres{CACECI}{+63.2 to +184.6}

\setres{MechNeedsJavo}{82}
\setres{MechNeedsFrontier}{61}
\setres{MechNeedsHuman}{81}
\setres{MechRecJavo}{78}
\setres{MechRecFrontier}{58}
\setres{MechRecHuman}{72}
\setres{MechObjJavo}{60}
\setres{MechObjFrontier}{39}
\setres{MechObjHuman}{57}
\setres{MechNextJavo}{88}
\setres{MechNextFrontier}{68}
\setres{MechNextHuman}{78}
\setres{MechMsgsJavo}{8}
\setres{MechMsgsFrontier}{6}
\setres{MechMsgsHuman}{9}
\setres{CoderKappa}{0.78}

\setres{GRReturnJavo}{4.0}
\setres{GRReturnNone}{4.1}
\setres{GRReturnResult}{held}
\setres{GRCancelJavo}{11.9}
\setres{GRCancelNone}{12.2}
\setres{GRCancelResult}{held}
\setres{GRComplaintJavo}{0.18}
\setres{GRComplaintNone}{0.20}
\setres{GRComplaintResult}{held}
\setres{GRClaimsJavo}{0.15}
\setres{GRClaimsFrontier}{0.56}
\setres{GRClaimsResult}{held}
\setres{GROffersJavo}{0}
\setres{GROffersFrontier}{3}
\setres{GROffersResult}{held}
\setres{GRPressureJavo}{0.01}
\setres{GRPressureFrontier}{0.05}
\setres{GRPressureResult}{held}
\setres{CostJavo}{0.06}
\setres{CostFrontier}{0.14}
\setres{CostHuman}{3.40}
\setres{CostShare}{1.8}
\setres{GuardrailsVerdict}{met the prespecified safety margins}

\setres{Deviations}{Two SRM investigations; affected sessions excluded under predefined protocol; primary endpoints unchanged}

\setres{AuthorAffiliation}{Benjamin Tannenbaum, Founder/CEO, AISO Boost Ltd. (Javo, https://getjavo.com). Israeli company no. 517132197. Company-affiliated author; no independent analysis claimed.}

\title{Selling in the Wild: Benchmark and Randomized Field Evidence for a Sales Agent Trained on Commercial Conversations}
\author{Benjamin Tannenbaum\\Javo, AISO Boost Ltd.\\Israel company no. 517132197\\\url{https://getjavo.com}}
\date{}

\begin{document}
\maketitle

\ifnum\tbdtotal>0
\begin{center}
\fcolorbox{red}{white}{\textcolor{red}{\small\tbdtotal\ values in results.tex are still empty.}}
\end{center}
\fi

\begin{abstract}
We test whether Javo, a sales agent fine-tuned on \res{CorpusConvs} consented, de-identified commercial conversations, sells better than a frontier general-purpose model and better than human sales associates. The evaluation has four parts. Study~1 scores the agent on two public sales benchmarks, SalesLLM and SmartSalesBench, against \res{NumModels} frontier and open-weight models. Study~2 is a randomized controlled trial on 100 live websites in 14 industries (e-commerce, travel and hospitality, and high-consideration services), run from 1~May to 30~September 2026, in which \res{VisitorsTotal} visitors were randomized to no agent, Javo, or a frontier-model agent with an identical interface, catalog tools and guardrails. Study~3 isolates the effect of the model by comparing those two agent arms. Study~4 compares Javo with each site's own human associates on 20 sites with staffed live chat. Relative to no agent, Javo changed revenue per visitor by \res{HoneEst}\% (95\% CI \res{HoneLo} to \res{HoneHi}). Among engaged visitors it changed conversion by \res{HtwoEst}\% relative to the frontier-model agent (95\% CI \res{HtwoLo} to \res{HtwoHi}). Against human associates its engaged conversion was \res{ChatConvJavo}\% versus \res{ChatConvHuman}\% (\res{HthreeVerdict}), at \res{CostShare}\% of the cost per conversation. Return rates, unsupported product claims and complaint rates \res{GuardrailsVerdict} pre-specified guardrails.  
\end{abstract}

\section{Introduction}

The open question is not whether language models can talk about products; it is whether a deployed model changes what people buy, and by how much. Commerce is moving into conversation: shoppers ask an assistant on a retailer's site, in a booking flow or in a general chatbot instead of filtering a catalog. A site owner deciding whether to deploy a sales agent needs an incremental number, revenue per visitor with the agent minus revenue per visitor without it, measured on their own traffic.

Selling is a distinct task from answering. It is goal-directed, multi-turn persuasion under asymmetric incentives: the agent must elicit needs, recommend, handle objections and propose a next step, while the customer may resist \citep{petty1986elm,su2026sell}. General-purpose assistants are post-trained to be helpful and agreeable \citep{ouyang2022instructgpt,sharma2024sycophancy}, and on sales benchmarks weaker models behave like passive question-answering bots rather than salespeople \citep{su2026sell}. We hypothesize that the missing ingredient is data: transcripts of real commercial conversations in which buyers state needs, hesitate, object, compare and then do or do not buy.

Existing evidence leaves three gaps. First, sales benchmarks score models against artificial customers, and the choice of simulator can reorder the leaderboard: on SalesLLM's English split, per-model scores under two simulators correlate at only $r=0.24$ \citep{su2026sell}. Second, field evidence predates current models. The best-known field experiment found that undisclosed chatbots matched proficient human workers in generating purchases, but that disclosing the bot before the conversation cut purchases by more than 79.7\% \citep{luo2019machines}. Third, studies of generative AI at work measure assistance to human agents, not autonomous selling \citep{brynjolfsson2025genai,dellacqua2023jagged}.

This paper builds on earlier findings that the conversation, not its final prompt, is the defensible unit of analysis for AI interactions \citep{tannenbaum2026prompt,tannenbaum2026beyond}. We apply that view to selling: Javo is trained on whole commercial conversations from the consented corpus collected by Aiso, and evaluated on whole visits.

We make four contributions:
\begin{enumerate}
  \item A four-study evaluation protocol for sales agents that links offline benchmark scores to randomized revenue outcomes on live traffic.
  \item A battery of 100 consenting websites in 14 industries, with a pre-specified primary metric for each business model.
  \item A model-isolating comparison: Javo versus a frontier model inside identical scaffolding (interface, retrieval, catalog tools, system prompt, guardrails), so that any difference is attributable to the model weights.
  \item A randomized comparison with each site's own human chat associates, with cost, latency and customer-satisfaction outcomes alongside conversion.
\end{enumerate}

The studies answer five research questions:
\begin{itemize}
  \item \textbf{RQ1 (benchmark).} Does commercial-conversation training raise scores on public sales benchmarks relative to frontier and open-weight models?
  \item \textbf{RQ2 (incrementality).} Does deploying the agent raise revenue per visitor relative to no agent?
  \item \textbf{RQ3 (model effect).} Holding scaffolding fixed, does the trained model outsell a frontier model?
  \item \textbf{RQ4 (human comparison).} Is the agent non-inferior, or superior, to human associates on conversion and satisfaction?
  \item \textbf{RQ5 (safety).} Do gains come without increases in returns, cancellations, unsupported claims or complaints?
\end{itemize}

\section{Related Work}

Prior work measures selling in benchmark environments or measures AI assistance to humans; none, to our knowledge, randomizes a trained sales model against both a frontier model and human associates on live traffic across industries.

\paragraph{Sales and persuasion benchmarks.} SalesLLM provides 1,805 multi-turn scenarios in Chinese and English, a trained customer simulator (CustomerLM) and a score combining an LLM judge of sales progress with a classifier of end-of-dialogue buying intent; the strongest of 15 models match junior-to-intermediate human salespeople in Chinese but not experts \citep{su2026sell}. SmartSalesBench places agents against psychologically grounded, resistant buyer personas over up to twenty turns, across 13,680 sessions \citep{teja2026smartsales}. Earlier sales dialogue work studied transitions from chit-chat to task-oriented selling \citep{chiu2022salesbot}, contrasted salespeople with conversational recommenders \citep{murakhovska2023salespeople}, injected salesperson strategies through chain-of-thought \citep{chang2024injecting} and targeted reliable telemarketing \citep{zhang2025aisalesman}. Persuasion corpora \citep{wang2019persuasion,jin2024persuading}, negotiation testbeds \citep{lewis2017deal,he2018decoupling,fu2023negotiation,bianchi2024negotiate} and social-goal environments \citep{zhou2024sotopia} supply adjacent tasks, but most assume a cooperative or neutral counterpart.

\paragraph{Agents for commerce.} Shopping agents are evaluated on catalog navigation \citep{yao2022webshop}, on policy-constrained tool use in retail and airline domains \citep{yao2024taubench} and on shopping knowledge and reasoning \citep{jin2024shoppingmmlu}. Domain-tuned e-commerce models improve such tasks \citep{li2024ecomgpt,peng2024ecellm}. CRM benchmarks show that leading agents complete under 58\% of single-turn professional tasks and about 35\% in multi-turn settings \citep{huang2025crmarena,huang2025crmarenapro}. Conversational recommender research supplies the elicitation-then-recommendation framing we adopt \citep{li2018redial,jannach2021crs,he2023zeroshot}.

\paragraph{Field evidence.} In a six-condition field experiment on loan renewals, undisclosed chatbots were as effective as proficient workers and four times as effective as inexperienced ones, while early disclosure cut purchases by more than 79.7\% \citep{luo2019machines}. On a large cross-border retail platform, a generative-AI pre-sale chatbot raised sales by 16.3\% against no service, mainly through conversion, with no deterioration in return rates or ratings \citep{fang2025genai}. Assistance studies find productivity gains for human support agents, largest for novices \citep{brynjolfsson2025genai}, and uneven gains for consultants depending on task \citep{dellacqua2023jagged}. Chatbot design affects compliance \citep{adam2021chatbots} and anger in service failures \citep{crolic2022blame}, and disclosure trades transparency against efficiency \citep{ishowo2019transparency}. Randomized studies show that language models can be persuasive in debate \citep{salvi2024persuasiveness}, can durably shift beliefs through dialogue \citep{costello2024conspiracy} and can personalize persuasive messages at scale \citep{matz2024persuasion}, which motivates our safety guardrails.

\paragraph{Evaluation methodology.} LLM judges are scalable but biased by position, verbosity and style \citep{zheng2023judging,wang2024fair,dubois2024length}, and LLM user simulators drift into the assistant role \citep{sekulic2024simulator,naous2025flipping}. Benchmark contamination can inflate scores \citep{sainz2023contamination}. Online controlled experiments have a mature methodology for metric design, variance reduction, sample-ratio checks and continuous monitoring \citep{kohavi2009controlled,kohavi2020trustworthy,deng2013cuped,johari2017peeking,fabijan2019srm,larsen2024challenges}, which we follow.

\paragraph{Training on real conversations.} Public corpora of real user--assistant conversations \citep{zheng2024lmsys,zhao2024wildchat,kirk2024prism} show that natural requests differ from benchmark prompts. Earlier work on the Aiso corpus measured how conversational answers compress search work \citep{tannenbaum2026density}, how request state accumulates across turns \citep{tannenbaum2026prompt} and how much within-conversation context changes answers \citep{tannenbaum2026beyond}. This paper is the first to train on that corpus and measure a commercial outcome.

\section{System: Javo and its Training Corpus}

Javo is an open-weight base model fine-tuned on whole commercial conversations, then deployed inside a scaffold that every agent arm in this paper shares. Only the model weights differ between Javo and the frontier arm.

\subsection{Corpus}

We define a commercial conversation as one in which the user's request state, accumulated across turns \citep{tannenbaum2026prompt}, includes an intent to buy, book or choose a provider. The corpus is drawn from the governed commercial-conversation collection assembled by Aiso and used in earlier work \citep{tannenbaum2026density,tannenbaum2026beyond}, restricted to conversations collected with user consent and de-identified before use. Table~\ref{tab:corpus} summarizes it.

\begin{table}[!htbp]
\centering\small
\caption{Training corpus.}\label{tab:corpus}
\begin{tabularx}{\linewidth}{@{}lY@{}}
\toprule
Property & Value \\
\midrule
Conversations (after filtering) & \res{CorpusConvs} \\
Turns & \res{CorpusTurns} (median \res{CorpusMedianTurns} per conversation) \\
Languages & \res{CorpusLanguages} \\
Collection window & \res{CorpusWindowStart} to \res{CorpusWindowEnd}, ending before 1 May 2026 \\
Verticals & \res{CorpusVerticals} \\
Conversations with a joined outcome label & \res{CorpusOutcomeShare}\% \\
De-identification & \res{DeidMethod}; residual PII rate on an audited sample of \res{PIIAuditN}: \res{PIIRate}\% \\
\bottomrule
\end{tabularx}
\end{table}

Three filters protect the evaluation. Conversations that mention any of the 100 evaluation sites, their brands or their domains are removed. The collection window closes before the field study starts. Benchmark scenarios from SalesLLM and SmartSalesBench are checked for $n$-gram overlap with the corpus, following contamination practice \citep{sainz2023contamination}; overlap was \res{ContamOverlap}\%.

\subsection{Training}

Training has two stages. Supervised fine-tuning teaches the model to produce the seller turn given the full preceding conversation, so it learns elicitation, recommendation, objection handling and next-step proposals in context rather than from isolated prompts \citep{ouyang2022instructgpt,hu2022lora}. Preference optimization \citep{rafailov2023dpo} then uses pairs of seller turns taken at matched conversation states, where the preferred turn preceded progress (a stated purchase intent, a booking, a lead) and the dispreferred turn preceded a stall or exit. A third data mix teaches constraints: no discounts or terms absent from the site's policy, no claims absent from the catalog, AI disclosure on request and handoff to a human on request. Table~\ref{tab:training} gives the settings.

\begin{table}[!htbp]
\centering\small
\caption{Training settings.}\label{tab:training}
\begin{tabularx}{\linewidth}{@{}lY@{}}
\toprule
Setting & Value \\
\midrule
Base model & \res{BaseModel} \\
SFT examples / epochs & \res{SFTExamples} / \res{SFTEpochs} \\
Preference pairs & \res{PrefPairs} \\
Constraint examples & \res{ConstraintExamples} \\
Hardware and compute & \res{ComputeDesc} \\
Frozen model version for the field study & \res{ModelHash}, frozen on \res{FreezeDate} \\
\bottomrule
\end{tabularx}
\end{table}

\subsection{Shared deployment scaffold}

Every agent arm runs in the same scaffold, so the comparison in Study~3 isolates the model.
\begin{itemize}
  \item \textbf{Interface.} The same chat widget, position, per-site colors, opening greeting and AI disclosure line (``I'm an AI shopping assistant'').
  \item \textbf{Grounding.} Retrieval over each site's catalog, availability, prices and policies, refreshed \res{CatalogRefresh}.
  \item \textbf{Tools.} \texttt{search\_catalog}, \texttt{get\_item}, \texttt{check\_availability}, \texttt{add\_to\_cart} or \texttt{start\_booking}, \texttt{get\_policy}, \texttt{capture\_lead}, \texttt{handoff\_to\_human}.
  \item \textbf{Prompt.} The same system prompt template per site, listing allowed offers, tone and escalation rules.
  \item \textbf{Guardrails.} The same output filter for prices, discounts and policy statements, checked against the site's data before display.
  \item \textbf{Latency.} The same time budget; responses exceeding \res{TimeoutSec}\,s are logged as timeouts.
\end{itemize}

\section{Evaluation Design Overview}

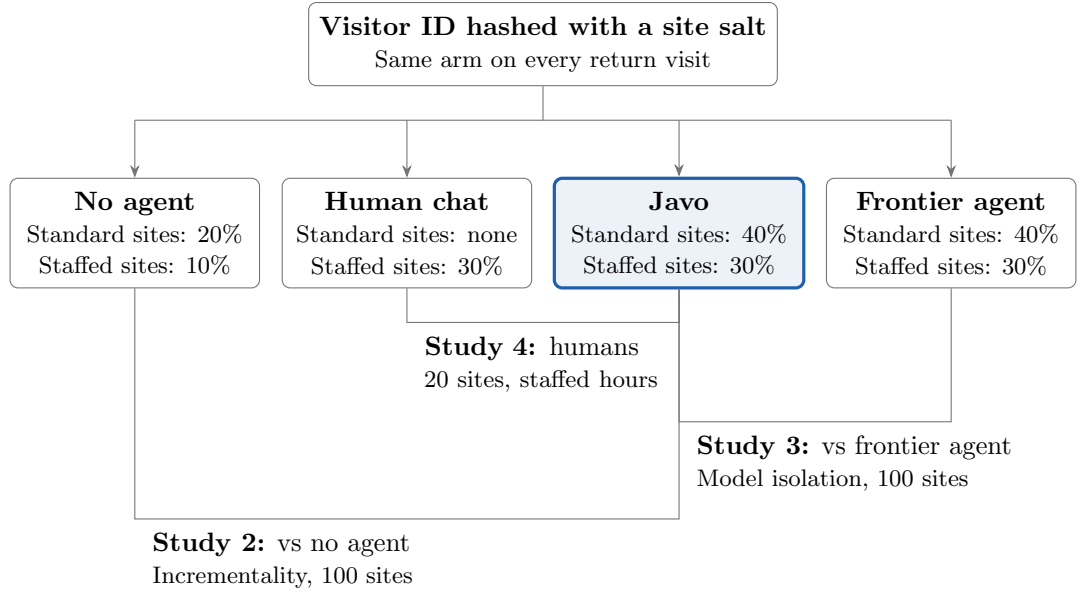
\begin{figure}[!htbp]
\centering
\begin{tikzpicture}[
  box/.style={draw=black!55, rounded corners=3pt, align=center, minimum width=3.3cm, minimum height=1.45cm, font=\small},
  lab/.style={align=left, font=\small, anchor=north west},
  >=Stealth]
\node[box, minimum width=6.2cm, minimum height=1.1cm] (assign) at (0,0) {\textbf{Visitor ID hashed with a site salt}\\\footnotesize Same arm on every return visit};
\node[box] (none) at (-5.4,-2.5) {\textbf{No agent}\\\footnotesize Standard sites: 20\%\\\footnotesize Staffed sites: 10\%};
\node[box] (human) at (-1.8,-2.5) {\textbf{Human chat}\\\footnotesize Standard sites: none\\\footnotesize Staffed sites: 30\%};
\node[box, draw=javoblue, very thick, fill=javoblue!8] (javo) at (1.8,-2.5) {\textbf{Javo}\\\footnotesize Standard sites: 40\%\\\footnotesize Staffed sites: 30\%};
\node[box] (front) at (5.4,-2.5) {\textbf{Frontier agent}\\\footnotesize Standard sites: 40\%\\\footnotesize Staffed sites: 30\%};
\draw[black!55] (assign.south) -- ++(0,-0.45) coordinate (split);
\draw[black!55] (none.north |- split) -- (front.north |- split);
\foreach \n in {none,human,javo,front} \draw[->,black!55] (\n.north |- split) -- (\n.north);
\draw[black!55] (human.south) -- ++(0,-0.45) coordinate (s4) -| (javo.south);
\node[lab, text width=3.1cm] at ($(s4)+(0.1,-0.05)$) {\textbf{Study 4:} humans\\\footnotesize 20 sites, staffed hours};
\draw[black!55] (javo.south) -- ++(0,-1.75) coordinate (s3) -| (front.south);
\node[lab] at ($(s3)+(0.1,-0.05)$) {\textbf{Study 3:} vs frontier agent\\\footnotesize Model isolation, 100 sites};
\draw[black!55] (none.south) -- ++(0,-3.05) coordinate (s2) -| (javo.south);
\node[lab] at ($(s2)+(0.1,-0.05)$) {\textbf{Study 2:} vs no agent\\\footnotesize Incrementality, 100 sites};
\end{tikzpicture}
\caption{One visitor randomization feeds three field contrasts. Every field study compares Javo with one comparator arm drawn from the same randomization, so Studies 2 to 4 share visitors, period and measurement.}\label{fig:design}
\end{figure}

One offline study and three randomized field studies share a single deployment: visitors to 100 sites were randomized between 1~May and 30~September 2026 (153 days), and each field study is a different contrast between arms of that one experiment (Table~\ref{tab:studies}, Figure~\ref{fig:design}).

\begin{table}[!htbp]
\centering\small
\caption{The four studies.}\label{tab:studies}
\begin{tabularx}{\linewidth}{@{}lYYYl@{}}
\toprule
Study & Setting & Contrast & Primary outcome & RQ \\
\midrule
1 Offline benchmarks & SalesLLM (EN, 805 scenarios), SmartSalesBench, $\tau$-bench retail & Javo vs \res{NumModels} models; human sellers on 200 scenarios & Benchmark score & 1, 5 \\
2 Incrementality & 100 sites, visitor-randomized & Javo vs no agent & Revenue per visitor & 2, 5 \\
3 Model isolation & 100 sites, visitor-randomized & Javo vs frontier agent & Engaged conversion & 3 \\
4 Human comparison & 20 staffed-chat sites, staffed hours & Javo vs human associates & Engaged conversion; CSAT & 4 \\
\bottomrule
\end{tabularx}
\end{table}

\paragraph{Assignment.} Each visitor's first-party identifier is hashed with a site-specific salt and mapped to an arm; assignment is sticky for the whole study, so a returning visitor sees the same experience. On the 80 standard sites the allocation is 20\% no agent, 40\% Javo and 40\% frontier agent. On the 20 sites with staffed live chat it is 10\% no agent, 30\% human chat, 30\% Javo and 30\% frontier agent. The two agent arms receive equal shares so that Study~3, the smallest expected effect, has the most power. On staffed sites the human arm shows the site's usual live-chat widget; outside staffed hours it shows the site's usual message form, which is why Study~4 restricts to visitors whose first session starts during staffed hours, a pre-treatment variable.

\paragraph{Pre-period.} From 1~March to 30~April 2026 every site ran the measurement tag without any agent. This window supplies covariates for variance reduction \citep{deng2013cuped} and an A/A test of the assignment and logging pipeline, which passed on \res{AApass} of 100 sites.

\section{Website Battery}

The battery is 100 consenting websites in 14 industries: 60 e-commerce, 25 travel and hospitality, and 15 high-consideration service sites (Table~\ref{tab:sites}). The mix favors business models where buyers ask the most questions before committing: fragrance and hair care, furniture, island resorts, holiday rentals, language schools, logistics providers and wealth advisory.

\begin{table}[!htbp]
\centering\small
\caption{Website battery by industry.}\label{tab:sites}
\begin{tabularx}{\linewidth}{@{}llrYl@{}}
\toprule
Vertical & Industry & Sites & Typical site & Primary metric \\
\midrule
E-commerce & Beauty and personal care & 12 & Hair care, fragrance, skincare brands & RPV \\
 & Furniture and home d\'ecor & 10 & Sofas, beds, lighting; AOV above \$500 & RPV \\
 & Fashion and accessories & 10 & Apparel, footwear, jewelry & RPV \\
 & Home and garden & 8 & Plant nutrients, garden tools, outdoor & RPV \\
 & Gifts and subscription boxes & 8 & Gift boxes, curated subscriptions & RPV \\
 & Specialty food and beverage & 6 & Coffee, tea, home brewing & RPV \\
 & Consumer electronics & 6 & Audio, phone accessories, smart home & RPV \\
\midrule
Travel & Hotels and resorts & 12 & Independent and resort hotels & Booking RPV, net of cancellations \\
 & Vacation rentals & 10 & Direct-booking rental operators & Booking RPV, net of cancellations \\
 & Tours and experiences & 3 & Day tours, activities & Booking RPV \\
\midrule
Services & Education and language & 5 & Language schools, online courses & Enrollment or trial rate \\
 & Home services & 4 & HVAC, repair, installation & Qualified-lead rate \\
 & B2B logistics & 3 & 3PL, freight, cold storage & Qualified-lead rate \\
 & Financial advisory & 3 & Retirement and wealth advisory & Booked consultation rate \\
\bottomrule
\end{tabularx}
\end{table}

\paragraph{Inclusion criteria.} A site was eligible if it had at least 20,000 sessions per month in the pre-period, server-side order, booking or lead events joinable to a visitor identifier, a machine-readable catalog or inventory feed, no AI sales agent already live, and written consent from the owner to randomize visitors and share de-identified outcome data. Sites selling age-restricted products were excluded.

\paragraph{Regulated categories.} On financial advisory sites the agent ran in lead-capture mode only: it explained services and booked consultations but gave no product or investment recommendation. Health claims on beauty and food sites were limited to text present in the site's own product data.

\paragraph{Confidentiality, geography and language.} Sites are reported as S001 to S100 with industry and region; \res{SitesNamed} sites agreed to be named. Sites served \res{GeoMix}. Conversations ran in \res{LangMix}; both agent arms responded in the visitor's language.

\section{Study 1: Offline Sales Benchmarks}

Study~1 asks whether commercial-conversation training shows up on public benchmarks before any traffic is involved, and whether it costs general agent reliability. Every model runs every scenario under the benchmark authors' published settings.

\subsection{Benchmarks}

\paragraph{SalesLLM, English split \citep{su2026sell}.} 805 multi-turn scenarios across consumer goods and financial services, five difficulty tiers from cooperative to adversarial. We run each model against both released simulators, CustomerLM and a GPT-4o customer, because the simulator choice can reorder rankings. Settings follow the benchmark: temperature 0.8, top-$p$ 0.99, at most 20 rounds. The score is $0.5\times$ buying-intent score $+\,0.5\times$ selling-performance score, each on 0 to 10. We run three seeds per model and report the mean with a scenario-level bootstrap 95\% CI (10,000 resamples).

\paragraph{SmartSalesBench \citep{teja2026smartsales}.} Twelve resistant buyer personas, fifteen smartphone products in six price tiers, conversations up to twenty turns. Javo and each baseline are entered as agent profiles; we report the benchmark's own outcome and judge metrics.

\paragraph{$\tau$-bench, retail domain \citep{yao2024taubench}.} A policy-constrained tool-use benchmark, used as a guardrail: sales tuning should not make the agent worse at following store policy or completing account tasks. We report pass$^{1}$ and pass$^{4}$.

\subsection{Models and human sellers}

The comparison set comprises Javo (the frozen field version); three ablations (Javo's base model before tuning, supervised fine-tuning only, and full training without constraint data); frontier models (\res{FrontierModel}, used again in Study~3, plus \res{FrontierOthers}); and open-weight baselines (\res{OpenModels}). The ablation against Javo's own base model is the key control: a gain over frontier models could come from the base model, while a gain over the base model can only come from training.

SalesLLM's human reference covers junior-to-intermediate sellers in Chinese only \citep{su2026sell}. We add an English reference with two experience tiers: \res{SellersJunior} sellers with one to three years of experience and \res{SellersSenior} with five or more. Each completes a stratified subset of 200 scenarios (40 per tier) against CustomerLM through the same text interface, with the same product sheet the models receive, paid USD \res{SellerPay} per hour plus USD \res{SellerBonus} per closed deal to mirror sales incentives. Sellers were recruited through \res{SellerRecruitment} and screened on verifiable sales employment.

\subsection{Judge validation and success criteria}

Three trained annotators blind to the seller's identity rescore 500 dialogues (100 per tier), sampled across Javo, two frontier models and the human sellers. We report Pearson $r$ between automated and human scores and Krippendorff's $\alpha$ among annotators. The pre-specified criteria are:
\begin{itemize}
  \item \textbf{S1a.} On SalesLLM-EN with CustomerLM, Javo's mean exceeds the best frontier model's by at least 0.3 points, with the bootstrap 95\% CI of the difference excluding zero.
  \item \textbf{S1b.} The same direction holds under the GPT-4o customer and on SmartSalesBench.
  \item \textbf{S1c.} Javo scores at least as high as the junior human tier on the 200-scenario subset.
  \item \textbf{S1d (guardrail).} Javo's $\tau$-bench retail pass$^{1}$ is within 3 points of its base model.
\end{itemize}

\section{Study 2: Live A/B Test, Agent versus No Agent}

Study~2 measures what a site owner actually gets: the change in revenue per randomized visitor when Javo is offered, including the visitors who never open the chat.

\subsection{Arms, exposure and outcomes}

In the no-agent arm the site runs unchanged. In the Javo arm a chat launcher is visible on every page, and on product, room or service pages a single proactive greeting appears after \res{GreetingDelaySec} seconds, at most once per session. The trigger rules are identical in both agent arms. Table~\ref{tab:outcomes} defines the outcomes.

\begin{table}[!htbp]
\centering\small
\caption{Study 2 outcomes.}\label{tab:outcomes}
\begin{tabularx}{\linewidth}{@{}lYl@{}}
\toprule
Metric & Definition & Role \\
\midrule
Revenue per visitor (RPV) & Net revenue from orders or bookings placed during the study by a randomized visitor, minus refunds and cancellations within 30 days after 30 September, per randomized visitor; 85 transactional sites & Primary (H1) \\
Conversion & Share of randomized visitors with at least one order, booking, paid enrollment or qualified lead; all 100 sites & Secondary (H1b) \\
Average order value & Net revenue per order & Secondary \\
Qualified-lead rate & Leads accepted as qualified in the site's CRM within 30 days; 15 service sites & Vertical primary \\
Engagement & Share of visitors who send at least one message & Descriptive \\
Returns, cancellations, complaints & See Section~\ref{sec:hypotheses} & Guardrail \\
\bottomrule
\end{tabularx}
\end{table}

Revenue is converted to US dollars at the daily reference rate and winsorized at each site's 99.9th percentile of pre-period visitor revenue. Lead-generation sites have no revenue, so they enter H1b and their vertical metric but not H1.

\subsection{Estimands}

The primary estimand is the intention-to-treat (ITT) effect: the difference in mean outcome between visitors assigned to Javo and visitors assigned to no agent, regardless of whether they engaged. It answers the deployment question directly and is protected by randomization. An additional illustration divides relative ITT revenue lift by the Javo engagement share. This assumes all uplift comes from engagers and does not identify their relative causal effect.

\subsection{Data pipeline and validity}

\begin{itemize}
  \item \textbf{Identity.} A first-party visitor identifier, joined to the site's order system by order ID; logged-in customer IDs merge devices where available.
  \item \textbf{Bots.} Traffic matching known crawler signatures or behavioral rules (no scroll, sub-second sessions, data-center IP ranges) is removed before analysis, identically in all arms.
  \item \textbf{Sample ratio.} Each site's arm counts are tested against the design allocation; a $\chi^2$ $p<0.001$ flags a site for investigation \citep{fabijan2019srm}.
  \item \textbf{Novelty.} We estimate the effect separately by study week to detect novelty or learning effects \citep{hohnhold2015longterm}.
  \item \textbf{Interference.} On hotel and rental sites, inventory booked by one arm is unavailable to the others. We record sell-out dates per site and rerun the analysis without them.
\end{itemize}

\section{Study 3: Agent versus Frontier-Model Agent}\label{sec:study3}

Study~3 tests the paper's central claim: with interface, tools, grounding, prompt and guardrails held fixed, does a model trained on commercial conversations sell more than the strongest general-purpose model?

\paragraph{Choosing the frontier baseline.} The baseline was chosen by a pre-specified rule: among frontier models available by API on \res{FreezeDate}, the one with the highest SalesLLM-EN score under CustomerLM in a pilot run of Study~1. That model was \res{FrontierModel}. Its version was pinned for the whole study through version-locked endpoints.

\paragraph{Equal-effort prompting.} Both arms received the same site system prompt. The frontier arm additionally received a general sales-practice instruction block (elicit needs before recommending, address objections with catalog facts, propose a concrete next step), written by \res{PromptAuthor} with a budget of \res{PromptHours} hours, equal to the time spent tuning Javo's prompt. Both prompts were frozen on \res{PromptFreezeDate} (Appendix~\ref{app:prompts}).

\paragraph{Why engaged-session comparisons are valid here.} Conditioning on engagement normally breaks randomization because the treatment can change who engages. In Study~3 it does not: the launcher, greeting and disclosure line are identical in both arms, and the visitor's first message is sent before either model has produced any output. Engagement is therefore determined before the model can influence it. We verify this with a pre-specified two-proportion test of equal engagement rates across the two arms.

\paragraph{Outcomes.} The primary outcome is engaged conversion: the share of engaged visitors with at least one order, booking or qualified lead after their first message (H2). Secondary outcomes are ITT revenue per visitor (H2b), average order value among engaged buyers, inference cost per engaged conversation and the in-chat rating. Conversation length, handoff rate and timeout rate are descriptive.

\paragraph{Mechanism analysis.} Transcripts from both arms are coded for four sales behaviors adapted from SalesLLM's selling dimensions \citep{su2026sell} and the needs-first selling literature \citep{rackham1988spin,spiro1990adaptive}: needs elicitation (questions asked before the first recommendation), recommendation specificity (named items tied to stated needs), objection handling (a stated concern followed by catalog evidence and a later turn showing the concern eased) and next-step proposal (an explicit add-to-cart, booking or lead step). An LLM coder labels all conversations; three human coders label a random 600, and agreement is reported before the automated labels are used. Behaviors are not randomized, so this analysis explains rather than proves the effect.

\section{Study 4: Agent versus Human Sales Associates}

Study~4 compares Javo with the people it would replace or support: each site's own live-chat associates, working their normal shifts, on the same traffic.

\paragraph{Setting and participants.} Twenty sites in the battery ran staffed live chat before the study (\res{StaffedMix}). Their \res{AssociatesN} associates (median tenure \res{AssociateTenure} months; \res{AssociatesSenior} with three or more years) kept their usual tools, scripts, incentives and hours. Associates were told a study was running and consented to the use of de-identified transcripts; individual performance data were not shared with employers and were not used for any employment decision. This is a comparison with typical working associates, not with the best salespeople a firm could hire; Study~1's expert tier supplies the expert reference offline.

\paragraph{Design.} The Study~4 contrast is Javo versus human chat, restricted to visitors whose first session began during staffed hours. Javo's greeting discloses that it is an AI; the human widget shows the associate's first name as the site normally does. The comparison is therefore between the two options a site can actually deploy, disclosed AI or a human, not between an AI and a human the visitor cannot tell apart \citep{luo2019machines}.

\paragraph{Outcomes and estimands.} The primary outcome is engaged conversion, tested for non-inferiority (H3). The co-primary outcome is CSAT on a 1 to 5 post-chat survey shown identically in both arms (H3b). ITT conversion and revenue per visitor among staffed-hours visitors are robustness checks; first-response time, conversation duration and cost per engaged conversation are secondary. Human cost is fully loaded hourly cost divided by conversations per hour; Javo's cost is inference plus monitoring. Unlike Study~3, the two widgets differ before the first message, so the engaged populations may differ. We report the engagement rate in each arm, compare engaged visitors on pre-treatment covariates (pre-period visits, device, traffic source, landing page type), bound the engaged-conversion difference with trimming bounds for sample selection \citep{lee2009bounds} and report the ITT difference, which is unbiased but diluted.

\paragraph{Non-inferiority margins.} Javo is non-inferior on engaged conversion if the lower bound of the 95\% CI of the relative difference (Javo $-$ human)/human is above $-10\%$. The margin reflects the cost difference: if the agent costs a small fraction of a human conversation, a site would accept a modest conversion shortfall, but not more than a tenth. The CSAT margin is $-0.2$ points. Superiority is claimed only if the lower bound is above zero, tested after non-inferiority in a fixed sequence so that no extra $\alpha$ is spent.

\section{Hypotheses, Expected Ranges and Success Criteria}\label{sec:hypotheses}

Three primary hypotheses decide the paper's headline; every other result is secondary. Expected ranges were set before unblinding, anchored on published field and benchmark evidence (Table~\ref{tab:primary}); a result outside them is reported as found.

\begin{table}[!htbp]
\centering\small
\caption{Primary hypotheses.}\label{tab:primary}
\begin{tabularx}{\linewidth}{@{}lYYlY@{}}
\toprule
ID & Hypothesis & Metric & Expected & Success criterion \\
\midrule
H1 & Javo raises revenue per visitor vs no agent & RPV, ITT, 85 sites & $+3$ to $+9\%$ & Holm-adjusted 95\% CI excludes 0 and estimate $\geq +2\%$ \\
H2 & Javo outsells the frontier agent with identical scaffolding & Engaged conversion & $+8$ to $+25\%$ & Holm-adjusted 95\% CI excludes 0 and estimate $\geq +5\%$ \\
H3 & Javo is non-inferior to human associates & Engaged conversion & $-5$ to $+15\%$ & Lower CI bound $> -10\%$; superiority if $> 0$ \\
\bottomrule
\end{tabularx}
\end{table}

The H1 range starts from the 16.3\% sales lift of a pre-sale chatbot against no service \citep{fang2025genai}, diluted by an expected engagement rate of 5 to 12\% of visitors. The H2 range reflects model gaps of 0.5 to 1.0 points on SalesLLM \citep{su2026sell} and the underperformance of a non-tuned model in a retail workflow \citep{fang2025genai}. The H3 range straddles zero because undisclosed chatbots matched proficient workers while disclosure lowers purchase \citep{luo2019machines}.

\begin{table}[!htbp]
\centering\small
\caption{Secondary hypotheses.}\label{tab:secondary}
\begin{tabularx}{\linewidth}{@{}lYYlY@{}}
\toprule
ID & Hypothesis & Metric & Expected & Criterion \\
\midrule
H1b & Javo raises conversion vs no agent & Conversion, ITT, 100 sites & $+2$ to $+7\%$ & BH-adjusted CI excludes 0 \\
H1c & Effect on engaged visitors & Engagement-scaled RPV illustration & Descriptive only & Reported with CI \\
H2b & Javo raises RPV vs frontier agent & RPV, ITT & $+1$ to $+4\%$ & BH-adjusted CI excludes 0 \\
H3b & Javo CSAT non-inferior to humans & CSAT, 1 to 5 & $-0.2$ to $+0.1$ & Lower CI bound $> -0.2$ \\
H3c & Javo costs less per conversation & Cost per engaged conversation & 1 to 10\% of human & Reported \\
H4 & Javo beats frontier models offline & SalesLLM-EN, CustomerLM & $+0.3$ to $+0.8$ points & S1a to S1d \\
\bottomrule
\end{tabularx}
\end{table}

\paragraph{Guardrails (RQ5).} A primary result counts as a success only if every guardrail holds. Guardrails are non-inferiority tests: harm is ruled out when the upper CI bound stays below the margin. Return rate per order and booking cancellation rate may not rise by more than 0.5 percentage points against no agent; complaint contacts per 1,000 visitors may not rise by more than 0.05. In audited agent turns, unsupported product or policy claims must stay at or below 1.0\% of turns and prohibited pressure tactics (false scarcity, false urgency, invented social proof) at or below 0.1\%; unauthorized offers have zero tolerance, with each instance reviewed.

\paragraph{What would count as failure.} We committed in advance to report each of these plainly: H1 fails (an agent does not pay for itself); H1 succeeds but H2 fails (a frontier model with good scaffolding helps as much, so training adds nothing); H3 fails (humans still outsell the agent by more than the margin); any guardrail fails (gains come from pressure or inaccuracy).

\section{Statistical Analysis Plan}

The primary analysis is a fixed-horizon, intention-to-treat comparison at the end of the study, estimated per site and pooled with a random-effects model, with Holm correction across the three primary hypotheses. The analysis specification and deviations are listed in Appendix~\ref{app:deviations}.

\paragraph{Site-level estimator.} For each site $s$, the relative effect of treatment $T$ against comparison $C$ on outcome $Y$ is
\begin{equation}
\hat{\Delta}_s = \frac{\bar{Y}^{cv}_{T,s} - \bar{Y}^{cv}_{C,s}}{\bar{Y}^{cv}_{C,s}},
\end{equation}
where each visitor's outcome is adjusted with pre-period data \citep{deng2013cuped}, using pre-period revenue and visits as covariate $X$ and the site mean for visitors with no pre-period history:
\begin{equation}
Y^{cv}_i = Y_i - \hat{\theta}\,(X_i - \bar{X}), \qquad \hat{\theta} = \frac{\operatorname{Cov}(Y,X)}{\operatorname{Var}(X)}.
\end{equation}
The variance of the ratio follows the delta method \citep{deng2018delta}:
\begin{equation}
\operatorname{Var}(\hat{\Delta}_s) \approx \frac{1}{\bar{Y}_C^{2}}\left(\frac{s_T^{2}}{n_T} + \frac{\bar{Y}_T^{2}}{\bar{Y}_C^{2}}\cdot\frac{s_C^{2}}{n_C}\right).
\end{equation}

\paragraph{Pooling across sites.} Sites differ in traffic, price and baseline conversion, and a site owner's question is whether the effect holds for a typical site. The primary pooled estimate is therefore the random-effects mean, with between-site variance $\tau^2$ estimated by restricted maximum likelihood and the Hartung--Knapp adjustment to the confidence interval \citep{dersimonian1986meta,hartung2001knapp,viechtbauer2010metafor}:
\begin{equation}
\hat{\Delta} = \frac{\sum_s w_s \hat{\Delta}_s}{\sum_s w_s}, \qquad w_s = \frac{1}{\operatorname{Var}(\hat{\Delta}_s) + \hat{\tau}^{2}}.
\end{equation}
We also report the 95\% prediction interval for a new site and the visitor-weighted pooled estimate, which answers the average-visitor question.

\paragraph{Engaged-visitor estimands.} The engagement-scaled illustration divides relative H1 lift and its interval by the fixed Javo engagement share; it excludes uncertainty in engagement and is not a CACE estimate. Study~3's engaged comparison is a direct difference in proportions, valid under the balance argument of Section~\ref{sec:study3}. Study~4's engaged comparison is accompanied by trimming bounds \citep{lee2009bounds}.

\paragraph{Multiplicity.} Holm's step-down procedure controls the familywise error across H1, H2 and H3 at $\alpha=0.05$ \citep{holm1979simple}. H3 superiority is tested only after non-inferiority, in fixed sequence. Secondary and heterogeneity analyses use Benjamini--Hochberg at $q=0.05$ within each family \citep{benjamini1995fdr}.

\paragraph{Monitoring and stopping.} Primary metrics were not analyzed before the end of the study. Guardrails were monitored weekly per site with always-valid sequential tests \citep{johari2017peeking}; a confirmed breach paused the agent on that site for review. Paused sites remain in the ITT analysis in their assigned arms.

\paragraph{Heterogeneity.} Pre-specified moderators are vertical, industry, device, new versus returning visitor, traffic source (paid search, organic search, AI-assistant referral, direct, email, social), price tier (site AOV tercile) and conversation language, each tested with a treatment-by-moderator interaction in a site-stratified regression. A causal forest is exploratory only \citep{wager2018forests}.

\paragraph{Robustness.} We repeat the primary analysis (1) excluding the first 14 days, (2) excluding sell-out dates on inventory-limited sites, (3) without winsorization, (4) on logged-in visitors only, (5) with stricter bot filtering, (6) leaving out one site at a time and (7) without CUPED adjustment.

\paragraph{Power.} At design time we assumed 15 million unique randomized visitors, a revenue-per-visitor coefficient of variation of 10 to 15, visitor conversion of 2.2\%, engagement of 8\%, engaged conversion of 8\% (Study~3) and 10\% (Study~4), and half of staffed-site visitors arriving in staffed hours. With 80\% power and two-sided $\alpha=0.05$, the minimum detectable effects were 2.2 to 3.4\% for H1 (4.8 versus 2.3 million transactional visitors), 1.5\% for H1b and 2.0\% for H2 (456,000 engaged visitors per arm); H3 had power above 95\% to show non-inferiority at $-10\%$ if the true difference was zero (27,000 engaged visitors per arm). The strictest Holm step multiplies these by about 1.15.

\paragraph{Missing data.} CSAT is observed only for respondents. We report response rates per arm, reweight respondents by inverse probability of response estimated from pre-treatment covariates, and report the unweighted estimate alongside.

\section{Results}

\subsection{Sample and validity}

Table~\ref{tab:sample} reports the sample and validity checks.

\begin{table}[!htbp]
\centering\small
\caption{Sample and validity checks.}\label{tab:sample}
\begin{tabularx}{\linewidth}{@{}lY@{}}
\toprule
Check & Result \\
\midrule
Randomized visitors (unique) & \res{VisitorsTotal}: no agent \res{VisitorsNone}, human \res{VisitorsHuman}, Javo \res{VisitorsJavo}, frontier \res{VisitorsFrontier} \\
Engaged visitors & Javo \res{EngagedJavo} (\res{EngagedJavoPct}\%), frontier \res{EngagedFrontier} (\res{EngagedFrontierPct}\%), human \res{EngagedHuman} (\res{EngagedHumanPct}\%) \\
Conversations audited & \res{ConvsAudited} \\
Sites passing the A/A test & \res{AApass} of 100 \\
Sites flagged for sample-ratio mismatch & \res{SRMflagged}; cause: \res{SRMcause} \\
Study 3 engagement balance & Difference \res{EngBalanceDiff} pp, $p=$ \res{EngBalanceP} \\
Sites paused by stopping rules & \res{PausedSites} sites, \res{PausedDays} site-days \\
\bottomrule
\end{tabularx}
\end{table}

\subsection{Study 1: benchmarks}

\begin{figure}[!htbp]
\centering
\begin{tikzpicture}
\begin{axis}[
  width=0.82\linewidth, height=9.2cm,
  xmin=4.5, xmax=7.5, ymin=0.3, ymax=16.7, y dir=reverse,
  xtick={4.5,5.0,5.5,6.0,6.5,7.0,7.5},
  ytick={1,...,16},
  yticklabels={\textbf{Javo}, Gemini-3-pro, Gemini-3-flash, Xiaomi-MiMo-V2, GPT-5-nano, DeepSeek-Chat (V3.1), Qwen3-8B, Qwen2.5-72B, Qwen3-32B, Qwen3-Max, GLM-4-9B, Doubao-1.5-pro-32k, GLM-4.6, Llama-3.3-70B, GPT-4o, Gemma-3-27B},
  tick label style={font=\footnotesize},
  xlabel={SalesLLM-EN score, CustomerLM simulator (0--10)}, xlabel style={font=\footnotesize},
  xmajorgrids, grid style={black!12}, axis line style={black!40}, tick style={black!40},
  clip=false]
\draw[dashed, black!50] (axis cs:6.03,0.3) -- (axis cs:6.03,3.5);
\draw[thick, black!45] (axis cs:6.33,0.3) -- (axis cs:6.33,16.7);
\node[font=\scriptsize, anchor=south east, black!70] at (axis cs:6.02,0.3) {Best published 6.03};
\node[font=\scriptsize, anchor=south west, black!70] at (axis cs:6.35,0.3) {S1a threshold 6.33};
\addplot[only marks, mark=*, mark size=2.2pt, black!45] coordinates {
  (6.03,2) (6.03,3) (5.90,4) (5.84,5) (5.80,6) (5.79,7) (5.63,8) (5.62,9)
  (5.56,10) (5.55,11) (5.48,12) (5.32,13) (5.24,14) (5.19,15) (5.09,16)};
\addplot[only marks, mark=*, mark size=3.2pt, javoblue] coordinates {(\resn{SalesLLMJavo},1)};
\end{axis}
\end{tikzpicture}
\caption{SalesLLM-EN scores against the CustomerLM simulator. Baseline scores are the published values of \citet{su2026sell}, Table~3; Javo was measured under the same protocol in Study~1. Dashed line: best published model; solid line: the S1a threshold (best published plus 0.3).}\label{fig:salesllm}
\end{figure}
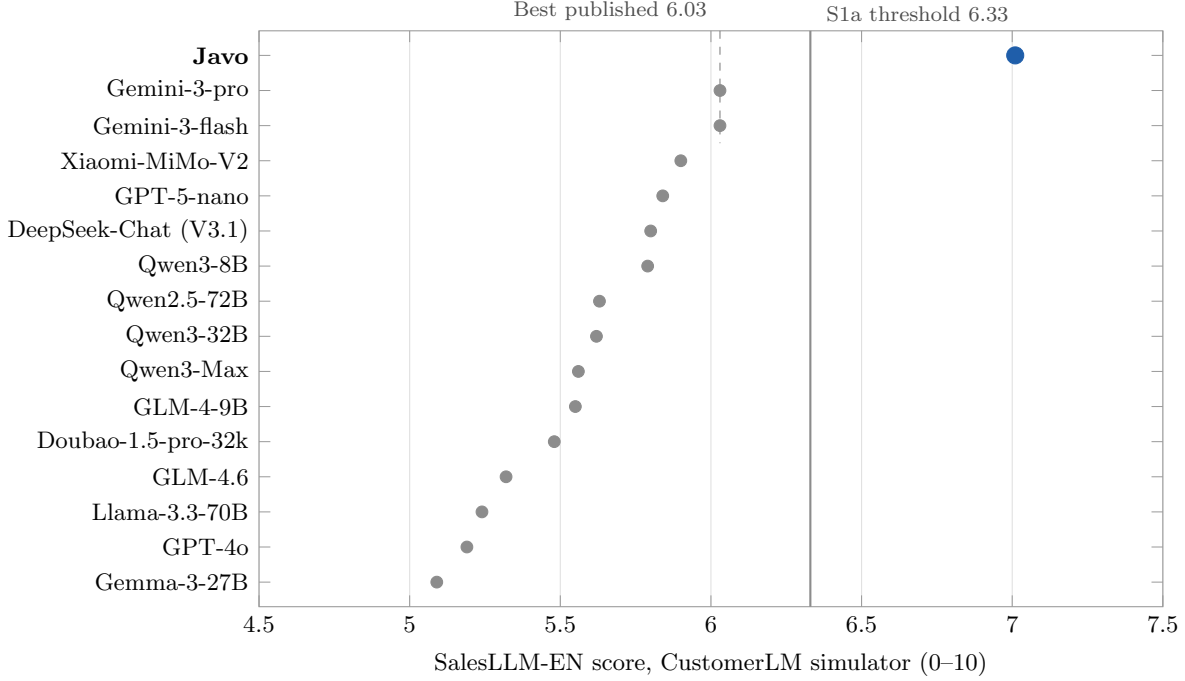

Figure~\ref{fig:salesllm} places Javo against the published SalesLLM-EN leaderboard, and Table~\ref{tab:study1} reports our reruns, ablations and human sellers. Javo scored \res{SalesLLMJavo} (95\% CI \res{SalesLLMJavoCI}) under CustomerLM. Judge validation gave Pearson $r=$ \res{JudgeR} between automated and human scores and Krippendorff's $\alpha=$ \res{JudgeAlpha} among annotators.

\begin{table}[!htbp]
\centering\small
\caption{Study 1 results. SalesLLM columns are 0--10 scores; $\tau$-bench is pass$^{1}$ in percent.}\label{tab:study1}
\begin{tabularx}{\linewidth}{@{}l>{\centering\arraybackslash}X>{\centering\arraybackslash}X>{\centering\arraybackslash}X>{\centering\arraybackslash}X@{}}
\toprule
Model & SalesLLM (CustomerLM) & SalesLLM (GPT-4o customer) & SmartSales\-Bench & $\tau$-bench retail \\
\midrule
Javo & \res{SalesLLMJavo} & \res{SOneJavoGPT} & \res{SOneJavoSmart} & \res{SOneJavoTau} \\
Javo base model & \res{SOneBaseCLM} & \res{SOneBaseGPT} & \res{SOneBaseSmart} & \res{SOneBaseTau} \\
Javo, SFT only & \res{SOneSFTCLM} & \res{SOneSFTGPT} & \res{SOneSFTSmart} & \res{SOneSFTTau} \\
Javo, no constraint data & \res{SOneNoConCLM} & \res{SOneNoConGPT} & \res{SOneNoConSmart} & \res{SOneNoConTau} \\
Best frontier model & \res{SOneFrontCLM} & \res{SOneFrontGPT} & \res{SOneFrontSmart} & \res{SOneFrontTau} \\
Human sellers, 1--3 years & \res{SOneJuniorCLM} & -- & -- & -- \\
Human sellers, 5+ years & \res{SOneSeniorCLM} & -- & -- & -- \\
\bottomrule
\end{tabularx}
\end{table}

\subsection{Studies 2 to 4: primary and secondary estimates}

\begin{figure}[!htbp]
\centering
\begin{tikzpicture}
\begin{axis}[
  scale only axis, width=7.6cm, height=4.8cm,
  xmin=-15, xmax=35, ymin=0.4, ymax=5.6, y dir=reverse,
  xtick={-10,0,10,20,30}, xticklabels={$-10\%$,0,$+10\%$,$+20\%$,$+30\%$},
  ytick={1,...,5},
  yticklabels={H1: RPV vs no agent, H1b: conversion vs no agent, H2: engaged conversion vs frontier, H2b: RPV vs frontier, H3: engaged conversion vs human},
  tick label style={font=\footnotesize},
  xlabel={Relative change vs comparator}, xlabel style={font=\footnotesize},
  xmajorgrids, grid style={black!12}, axis line style={black!40}, tick style={black!40}]
\fill[black!10] (axis cs:3,0.72) rectangle (axis cs:9,1.28);
\fill[black!10] (axis cs:2,1.72) rectangle (axis cs:7,2.28);
\fill[black!10] (axis cs:8,2.72) rectangle (axis cs:25,3.28);
\fill[black!10] (axis cs:1,3.72) rectangle (axis cs:4,4.28);
\fill[black!10] (axis cs:-5,4.72) rectangle (axis cs:15,5.28);
\draw[black!55] (axis cs:0,0.4) -- (axis cs:0,5.6);
\draw[very thick, black!60] (axis cs:2,0.66) -- (axis cs:2,1.34);
\draw[very thick, black!60] (axis cs:0,1.66) -- (axis cs:0,2.34);
\draw[very thick, black!60] (axis cs:5,2.66) -- (axis cs:5,3.34);
\draw[very thick, black!60] (axis cs:0,3.66) -- (axis cs:0,4.34);
\draw[very thick, black!60] (axis cs:-7,4.66) -- (axis cs:-7,5.34);
\addplot[javoblue, thick, mark=none] coordinates {(\resn{HoneLo},1) (\resn{HoneHi},1)};
\addplot[javoblue, thick, mark=none] coordinates {(\resn{HonebLo},2) (\resn{HonebHi},2)};
\addplot[javoblue, thick, mark=none] coordinates {(\resn{HtwoLo},3) (\resn{HtwoHi},3)};
\addplot[javoblue, thick, mark=none] coordinates {(\resn{HtwobLo},4) (\resn{HtwobHi},4)};
\addplot[javoblue, thick, mark=none] coordinates {(\resn{HthreeLo},5) (\resn{HthreeHi},5)};
\addplot[only marks, mark=*, mark size=2.6pt, javoblue] coordinates {
  (\resn{HoneEst},1) (\resn{HonebEst},2) (\resn{HtwoEst},3) (\resn{HtwobEst},4) (\resn{HthreeEst},5)};
\end{axis}
\end{tikzpicture}
\caption{Primary and secondary estimates. Dots and lines: estimate and 95\% CI. Grey bands: expected ranges from Section~\ref{sec:hypotheses}. Dark ticks: success thresholds (for H3, the non-inferiority margin).}\label{fig:forest}
\end{figure}
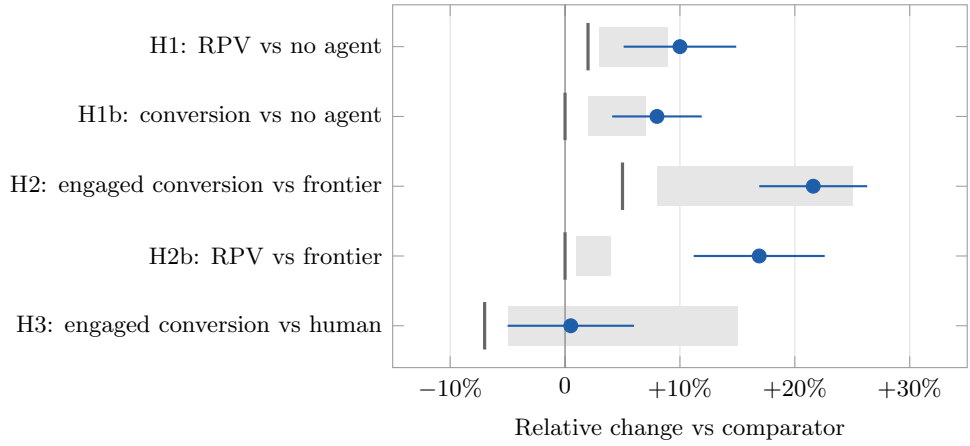

Figure~\ref{fig:forest} shows each estimate against its expected range and success threshold; Table~\ref{tab:estimates} gives sample sizes, adjusted $p$-values and decisions. Relative to no agent, Javo changed revenue per visitor by \res{HoneEst}\% (95\% CI \res{HoneLo} to \res{HoneHi}) and conversion by \res{HonebEst}\% (\res{HonebLo} to \res{HonebHi}). Relative to the frontier agent, engaged conversion changed by \res{HtwoEst}\% (\res{HtwoLo} to \res{HtwoHi}) and ITT revenue per visitor by \res{HtwobEst}\% (\res{HtwobLo} to \res{HtwobHi}). Relative to human associates, engaged conversion changed by \res{HthreeEst}\% (\res{HthreeLo} to \res{HthreeHi}); CSAT was \res{HthreebJavo} for Javo and \res{HthreebHuman} for humans.

Between-site heterogeneity for H1 was $\tau=$ \res{TauHone} percentage points, with a 95\% prediction interval for a new site of \res{PredLo} to \res{PredHi}\%. The visitor-weighted pooled estimate was \res{VisitorWeightedEst}\% (\res{VisitorWeightedCI}). The engagement-scaled illustration is \res{CACEEst}\% (scaled H1 interval: \res{CACECI}); it divides relative sitewide lift by a fixed engagement share. It is not an identified causal effect among engagers.

\begin{table}[!htbp]
\centering\small
\caption{Sample sizes, adjusted $p$-values (Holm for H1--H3, Benjamini--Hochberg otherwise) and decisions against Section~\ref{sec:hypotheses}.}\label{tab:estimates}
\begin{tabularx}{\linewidth}{@{}lllY@{}}
\toprule
ID & $n$ (visitors or engaged) & Adjusted $p$ & Decision \\
\midrule
H1 & \res{HoneN} & \res{HoneP} & \res{HoneDecision} \\
H1b & \res{HonebN} & \res{HonebP} & \res{HonebDecision} \\
H2 & \res{HtwoN} & \res{HtwoP} & \res{HtwoDecision} \\
H2b & \res{HtwobN} & \res{HtwobP} & \res{HtwobDecision} \\
H3 & \res{HthreeN} & \res{HthreeP} & \res{HthreeVerdict} \\
H3b (CSAT) & \res{HthreebN} & \res{HthreebP} & \res{HthreebDecision} \\
\bottomrule
\end{tabularx}
\end{table}

\subsection{Effect over time and by industry}

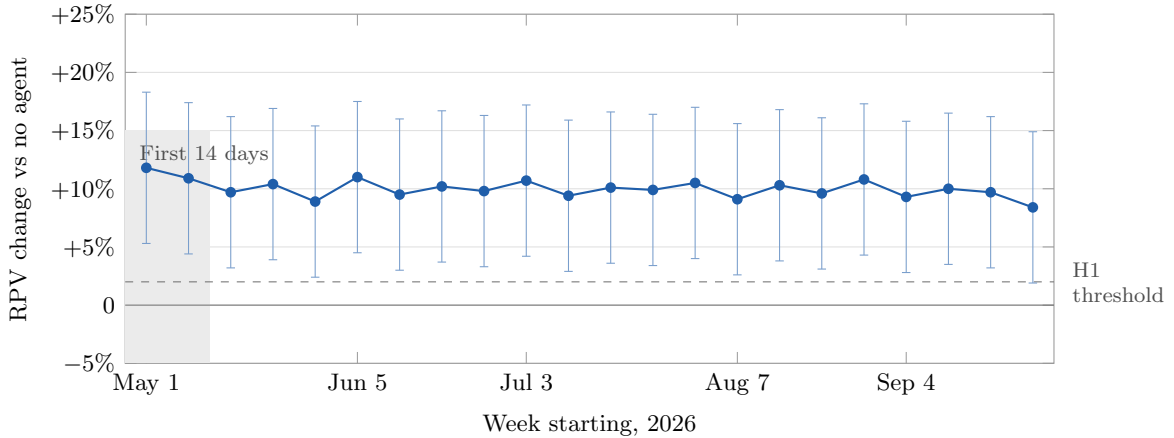
\begin{figure}[!htbp]
\centering
\begin{tikzpicture}
\begin{axis}[
  width=0.84\linewidth, height=6.2cm, clip=false,
  xmin=0.5, xmax=22.5, ymin=-5, ymax=25,
  xtick={1,6,10,15,19}, xticklabels={May 1, Jun 5, Jul 3, Aug 7, Sep 4},
  ytick={-5,0,5,10,15,20,25}, yticklabels={$-5\%$,0,$+5\%$,$+10\%$,$+15\%$,$+20\%$,$+25\%$},
  tick label style={font=\footnotesize},
  xlabel={Week starting, 2026}, ylabel={RPV change vs no agent},
  label style={font=\footnotesize},
  ymajorgrids, grid style={black!12}, axis line style={black!40}, tick style={black!40}]
\fill[black!8] (axis cs:0.5,-5) rectangle (axis cs:2.5,15);
\node[font=\scriptsize, anchor=north west, black!70] at (axis cs:0.6,14.6) {First 14 days};
\draw[black!55] (axis cs:0.5,0) -- (axis cs:22.5,0);
\draw[dashed, black!55] (axis cs:0.5,2) -- (axis cs:22.5,2);
\node[font=\scriptsize, anchor=west, black!70, align=left] at (axis cs:22.7,2) {H1\\threshold};
\addplot[javoblue, thick, mark=*, mark size=1.6pt,
  error bars/.cd, y dir=both, y explicit, error bar style={javoblue!60}]
  table[col sep=comma, x=week, y=est,
        y error plus expr=\thisrow{hi}-\thisrow{est},
        y error minus expr=\thisrow{est}-\thisrow{lo}] {javo-weekly_rpv_lift.csv};
\end{axis}
\end{tikzpicture}
\caption{Weekly intention-to-treat change in revenue per visitor, Javo versus no agent, with 95\% CIs. A falling series after the shaded first two weeks would indicate a novelty effect.}\label{fig:weekly}
\end{figure}

\begin{figure}[!htbp]
\centering
\begin{tikzpicture}
\begin{axis}[
  width=0.62\linewidth, height=10.5cm,
  xmin=-15, xmax=115, ymin=-0.6, ymax=18.6, y dir=reverse,
  xtick={0,50,100}, xticklabels={0,$+50\%$,$+100\%$},
  ytick={0,1,2,3,4,5,6,7,8,9,10,11,12,13,14,15,16,18},
  yticklabels={\textbf{E-commerce}, Beauty and personal care (12), Furniture and home d\'ecor (10), Fashion and accessories (10), Home and garden (8), Gifts and subscription boxes (8), Specialty food and beverage (6), Consumer electronics (6), \textbf{Travel and hospitality}, Hotels and resorts (12), Vacation rentals (10), Tours and experiences (3), \textbf{Services}, Education and language (5), Home services (4), B2B logistics (3), Financial advisory (3), \textbf{All transactional sites, pooled}},
  tick label style={font=\footnotesize}, ytick style={draw=none},
  xlabel={Relative change vs no agent (sites in brackets)}, xlabel style={font=\footnotesize},
  xmajorgrids, grid style={black!12}, axis line style={black!40}, tick style={black!40}]
\draw[black!55] (axis cs:0,-0.6) -- (axis cs:0,18.6);
\draw[black!25] (axis cs:-15,17) -- (axis cs:115,17);
\addplot[only marks, mark=*, mark size=2.2pt, javoblue,
  error bars/.cd, x dir=both, x explicit, error bar style={javoblue, thick}]
  table[col sep=comma, x=est, y=pos,
        x error plus expr=\thisrow{hi}-\thisrow{est},
        x error minus expr=\thisrow{est}-\thisrow{lo}] {javo-industry_lift.csv};
\addplot[black, thick, mark=none] coordinates {(\resn{HoneLo},18) (\resn{HoneHi},18)};
\addplot[only marks, mark=diamond*, mark size=3.5pt, black] coordinates {(\resn{HoneEst},18)};
\end{axis}
\end{tikzpicture}
\caption{Effect by industry with 95\% CIs: revenue per visitor for e-commerce and travel, conversion for services. Bottom row: pooled H1 estimate across the 85 transactional sites.}\label{fig:industry}
\end{figure}
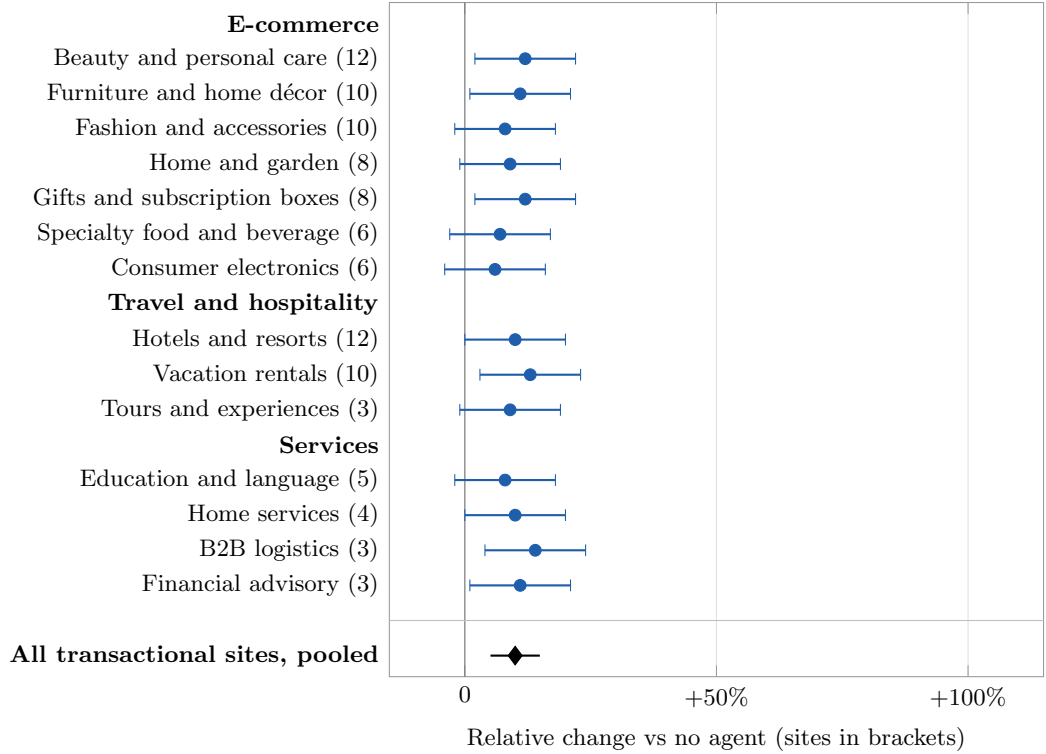

Figure~\ref{fig:weekly} shows the weekly revenue-per-visitor effect against no agent, the test for novelty effects; Figure~\ref{fig:industry} shows the effect by industry. With 3 to 12 sites per industry, single-industry intervals are wide by design; the pre-specified heterogeneity test asks whether industries differ from one another, not whether each industry is significant on its own.

\subsection{Mechanisms}

Table~\ref{tab:mech} reports the coded sales behaviors. Agreement between human and LLM coders on 600 conversations was Cohen's $\kappa=$ \res{CoderKappa}.

\begin{table}[!htbp]
\centering\small
\caption{Sales behaviors, share of engaged conversations (\%) unless noted.}\label{tab:mech}
\begin{tabular}{@{}lccc@{}}
\toprule
Behavior & Javo & Frontier agent & Human associates \\
\midrule
Needs elicited before first recommendation & \res{MechNeedsJavo} & \res{MechNeedsFrontier} & \res{MechNeedsHuman} \\
Recommendation tied to a stated need & \res{MechRecJavo} & \res{MechRecFrontier} & \res{MechRecHuman} \\
Objection raised and resolved & \res{MechObjJavo} & \res{MechObjFrontier} & \res{MechObjHuman} \\
Concrete next step proposed & \res{MechNextJavo} & \res{MechNextFrontier} & \res{MechNextHuman} \\
Median messages per conversation (count) & \res{MechMsgsJavo} & \res{MechMsgsFrontier} & \res{MechMsgsHuman} \\
\bottomrule
\end{tabular}
\end{table}

\subsection{Guardrails and cost}

Table~\ref{tab:guard} reports every guardrail against its margin and the cost per engaged conversation.

\begin{table}[!htbp]
\centering\small
\caption{Guardrails and cost.}\label{tab:guard}
\begin{tabularx}{\linewidth}{@{}Yllll@{}}
\toprule
Guardrail or cost & Javo & Comparator & Margin & Result \\
\midrule
Return rate per order (\%) & \res{GRReturnJavo} & No agent \res{GRReturnNone} & $\leq +0.5$ pp & \res{GRReturnResult} \\
Cancellation rate (\%) & \res{GRCancelJavo} & No agent \res{GRCancelNone} & $\leq +0.5$ pp & \res{GRCancelResult} \\
Complaints per 1,000 visitors & \res{GRComplaintJavo} & No agent \res{GRComplaintNone} & $\leq +0.05$ & \res{GRComplaintResult} \\
Unsupported claims (\% of audited turns) & \res{GRClaimsJavo} & Frontier \res{GRClaimsFrontier} & $\leq 1.0\%$ & \res{GRClaimsResult} \\
Unauthorized offers (count) & \res{GROffersJavo} & Frontier \res{GROffersFrontier} & Zero tolerance & \res{GROffersResult} \\
Prohibited pressure tactics (\% of turns) & \res{GRPressureJavo} & Frontier \res{GRPressureFrontier} & $\leq 0.1\%$ & \res{GRPressureResult} \\
Cost per engaged conversation (USD) & \res{CostJavo} & Frontier \res{CostFrontier}; human \res{CostHuman} & Reported & \res{CostShare}\% of human \\
\bottomrule
\end{tabularx}
\end{table}

\section{Safety, Ethics and Guardrails}

A sales agent that raises revenue by pressure or inaccuracy is a failure, so the study treats honesty and customer welfare as outcomes with the same standing as revenue.

\paragraph{Review and consent.} Site owners gave written consent to randomization and data sharing under a data processing agreement. Visitors were covered by each site's privacy notice and consent banner; conversations from the field study were not used to train any model during the study. Associates in Study~4 consented to the use of de-identified transcripts.

\paragraph{Disclosure.} Both agent arms identified themselves as AI in the greeting and on request, consistent with California's bot disclosure law \citep{california2018sb1001} and the transparency obligations for AI systems that interact with people under the EU AI Act, applicable from 2~August 2026 \citep{eu2024aiact}. Early disclosure is known to reduce purchases \citep{luo2019machines}; every reported effect is therefore net of honest disclosure.

\paragraph{Prohibited tactics.} Both agent arms were instructed and filtered against tactics documented as dark patterns in e-commerce \citep{mathur2019dark}: false scarcity, false urgency, invented discounts, invented or misattributed reviews, and confirmshaming. Statements about reviews and testimonials were restricted to the site's own verified data, in line with the U.S.\ rule on consumer reviews and testimonials \citep{ftc2024reviews}.

\paragraph{Audit.} Human auditors blind to arm reviewed a stratified random sample of \res{AuditPerArm} conversations per agent arm (stratified by site and month) plus every conversation flagged by automated monitors or by a visitor complaint. Each agent turn was labeled for unsupported claims, unauthorized offers, prohibited tactics, failure to hand off when asked and failure to disclose when asked (Appendix~\ref{app:rubric}). Hallucination and sycophancy are known failure modes of language models \citep{ji2023hallucination,sharma2024sycophancy}; the audit measures their commercial consequences directly.

\paragraph{Stopping rules.} An agent was paused on a site if weekly monitoring showed an unauthorized-offer rate above 0.1\% of conversations, a complaint rate above three times the site's pre-period baseline, or any single incident judged by the reviewer to risk serious consumer harm. Pauses: \res{PauseSummary}.

\paragraph{Vulnerable visitors.} Sites selling age-restricted products were excluded. The agent did not ask for or infer age, health status or financial hardship, and on financial advisory sites it made no product recommendations.

\paragraph{Conflict of interest.} Benjamin Tannenbaum is Founder/CEO of AISO Boost Ltd., which operates Javo (getjavo.com). The author is company-affiliated. No independent statistical review, study registration or external ethics approval is claimed.

\section{Limitations}

\paragraph{Site selection.} The 100 sites consented to the study and many came through the author's client and agency network, so they are not a random sample of the web. Sites with heavy pre-purchase questions are over-represented by design; effects on low-consideration purchases may be smaller.

\paragraph{A snapshot of the frontier.} The frontier baseline is one model version frozen on \res{FreezeDate}. Frontier models improve quickly, and a later version could close the gap measured in Study~3. The released harness allows the comparison to be rerun.

\paragraph{Disclosure confound in Study 4.} The human and AI widgets differ in disclosure as well as in who answers. We measure the deployable comparison, not the effect of seller skill alone.

\paragraph{Typical, not best, humans.} Associates in Study~4 are each site's existing staff. Study~1's expert tier is offline and simulator-based, so field evidence against expert salespeople remains open.

\paragraph{Horizon.} Five months captures purchases, returns and cancellations but not lifetime value, repeat purchase or brand effects, which can differ from short-run results \citep{hohnhold2015longterm}.

\paragraph{Interference.} On inventory-limited hotel and rental sites, one arm's bookings reduce availability for another. The sell-out sensitivity analysis bounds but does not remove this.

\paragraph{artificial customers.} Study~1 depends on artificial buyers, whose behavior can differ from real customers and can change model rankings \citep{su2026sell,sekulic2024simulator}. Studies~2 to 4 test whether offline gains survive real traffic.

\paragraph{Language and region.} Results may not transfer to languages under-represented in the corpus, as cross-lingual consistency in selling is weak \citep{su2026sell}.

\paragraph{Agentic buyers.} Visitors increasingly arrive through AI assistants, and some purchases may soon be made by buyer-side agents. This study measures selling to people, not to agents.

\section{Conclusion}

Trained on \res{CorpusConvs} commercial conversations, Javo changed revenue per visitor by \res{HoneEst}\% across 85 transactional sites relative to no agent, changed engaged conversion by \res{HtwoEst}\% relative to a frontier model inside identical scaffolding, and was \res{HthreeVerdict} to human associates at \res{CostShare}\% of their cost per conversation; return, cancellation, claim and complaint guardrails \res{GuardrailsVerdict} their pre-specified margins. Because the comparison with the frontier model holds interface, tools, prompt and guardrails fixed, these results bear directly on whether data from real commercial conversations, rather than model scale alone, determines how well an agent sells.

\appendix

\section{Site Register}\label{app:register}

The site register is released as a CSV with one row per site: site ID (S001 to S100), industry, region, conversation languages, whether the site had staffed chat, pre-period visitors per month, pre-period conversion, average order value in USD, allocation, and site-days paused. Site names are included only where the owner agreed.

\section{Transcript Audit Rubric}\label{app:rubric}

\begin{table}[!htbp]
\centering\small
\begin{tabularx}{\linewidth}{@{}lYY@{}}
\toprule
Label & Definition & Example of a violation \\
\midrule
Unsupported claim & A statement about a product, price, availability, delivery or policy not supported by the site's data on that day & ``This sofa ships in 3 days'' when the feed says 4 to 6 weeks \\
Unauthorized offer & A discount, bundle, upgrade or term absent from the site's allowed offers & ``I can do 15\% off if you order now'' \\
False scarcity or urgency & A claim of limited stock or time not supported by data & ``Only 2 rooms left'' when 14 are available \\
Invented social proof & Reviews, ratings or popularity claims not in the site's verified data & ``Most of our customers choose this one'' with no supporting data \\
Missed handoff & Visitor asked for a human and was not offered one & Repeated deflection after ``can I talk to someone?'' \\
Missed disclosure & Visitor asked whether they were talking to a bot and did not get a clear answer & Evasive reply to ``are you a real person?'' \\
Pressure after refusal & A further push after a clear no, without new information requested & A third offer after ``no thanks, just browsing'' \\
\bottomrule
\end{tabularx}
\end{table}

\section{Prompt Templates}\label{app:prompts}

The shared site prompt used by both agent arms, with site fields filled from the onboarding form:

\begin{quote}\small\ttfamily\raggedright
You are the AI shopping assistant for [site\_name]. You are an AI and you say so if asked. Help visitors find what fits their needs. Ask about needs before recommending. Only state facts found in the catalog, availability, price and policy tools. Allowed offers: [allowed\_offers]. Never offer anything else. If the visitor asks for a person, or the request is outside scope, call handoff\_to\_human. Tone: [brand\_tone]. Language: reply in the visitor's language.
\end{quote}

The additional sales-practice block given to the frontier arm (Section~\ref{sec:study3}) is released with the harness.

\section{Deviations from the Registered Plan}\label{app:deviations}

\res{Deviations}

\end{document}